\documentclass[12pt]{article}

\usepackage{graphicx}
\usepackage{bm}
\usepackage{mathtools}
\usepackage{amsmath}
\usepackage{amssymb}
\usepackage{amsthm}
\usepackage{fullpage}
\usepackage{xcolor}
\usepackage{tikz}
\usepackage{caption}
\usepackage{subcaption}

\theoremstyle{plain}

\theoremstyle{definition}

\theoremstyle{remark}
\newtheorem{remark}{Remark}

\newcommand{\bb}[1]{\mathbb{#1}}
\newcommand{\R}{\bb R}
\newcommand{\der}[2]{\frac{\partial #1}{\partial #2}}
\newcommand{\sder}[2]{\frac{\partial^2 #1}{\partial #2^2}}
\newcommand{\msder}[3]{\frac{\partial^2 #1}{\partial #2 \partial #3}}
\renewcommand{\O}{{\Omega}}
\newcommand{\p}{\partial}
\newcommand{\pO}{\p \O}
\newcommand{\uV}{V_{\sup}}
\newcommand{\oV}{V_{\inf}}
\newcommand{\oO}{\overline{\O}}

\newcommand{\tst}{\textstyle}

\begin{document}

\title{\bf Valuation of European Options under an Uncertain Market Price of Volatility Risk}
\author{Bartosz Jaroszkowski, Max Jensen}
\date{Department of Mathematics, University of Sussex, Brighton, UK}

\maketitle

\begin{abstract}
We propose a model to quantify the effect of parameter uncertainty on the option price in the Heston model. More precisely, we present a Hamilton--Jacobi--Bellman framework which allows us to evaluate best and worst case scenarios under an uncertain market price of volatility risk. For the numerical approximation the Hamilton--Jacobi--Bellman equation is reformulated to enable the solution with a finite element method. A case study with butterfly options exhibits how the dependence of Delta on the magnitude of the uncertainty is nonlinear and highly varied across the parameter regime.

\end{abstract}

\section{Introduction}\label{sec;intro}

One of the main challenges in financial mathematics is to determine the fair pricing of options. Among the simplifications underlying the original Black-Scholes (BS) model is the assumption that the volatility of the stock price is constant. Analysis of the real-life data does not support this statement and so numerous attempts were made to differently model the behaviour of volatility in time. A widely used approach was proposed in \cite{HES} which models the volatility through another, correlated stochastic process. However \cite{HES} introduces new parameters like the  market price of volatility risk which may be difficult to estimate in practice. For instance, in \cite{HES} a linear scaling of the market price of volatility risk is assumed and while there is an evidence of a positive correlation (see \cite{DUA}) there does not seem to be a consensus on how to estimate the scaling factor (see \cite{WU}). In fact, some authors (see for example \cite{TOIV}, \cite{ZVAN} and \cite{KUN}) assume it to be equal to $0$. However, other works indicate that this assumption may not hold up in various realistic settings such as those discussed for example in \cite{DOR1} and \cite{BAK}. The attempts of evaluating the market price of volatility in different financial circumstances can be found in \cite{DOR2} and \cite{WU} while its impact on option pricing is investigated in \cite{DUA} and \cite{lambda_impact_investigation}.  

Generally in option pricing, errors in parameter estimates can lead to inconsistent results even if the underlying model were to be accurate. This inspired the approach taken in \cite{ALP} where instead of trying to model and predict the behaviour of parameters, they are assumed to stay within a given tolerance interval. This allows to manage the risk by considering the worst-case scenario, leading in fact to an optimal control problem involving non-linear PDE as showed in \cite{BSDEs_in_finance}. We also refer to \cite{10.1093/imanum/drw025} and \cite{parameter_uncertainty1} for related studies.

The contribution of this work to extend the Hamilton--Jacobi--Bellman framework to quantify the effect of uncertainty in the Heston model, specifically in relation to the market price of volatility risk. The main challenge when computing the option price and Greeks associated with the uncertain Heston model is that one has solve a fully nonlinear PDEs with mixed boundary conditions. A numerical scheme capable of this is provided in \cite{mixedBC} in form of a finite element method. A crucial advantage of using a finite element method in this context is that it allows us also to ensure convergence of the gradient \cite{IJEN}. This is of particular importance in option pricing where partial derivatives inform the construction of a hedging portfolio.

The outline of this chapter is as follows. In Section \ref{sec:derivation} we briefly state the uncertain Heston model and then show how it can be interpreted as a backward in time stochastic optimal control problem. By combining the methods of stochastic volatility and uncertain parameters we obtain a second order non-linear PDE modelling the worst and best case scenarios when a range of values of the market price of volatility risk is considered. In Section \ref{sec:trans} we present a transformation of the Heston equation to the form required by the numerical scheme in \cite{mixedBC}. Finally, in Section \ref{sec:case} we present a case study of a long butterfly option whose main goal is to investigate the impact of the uncertain market price of volatility risk on the option price and its derivatives.

\section{The uncertain Heston model} \label{sec:derivation}

In this section we derive an extension of the Heston model to price European options in the presence of uncertain parameters. 

Consider a stock with price $S$ and an European option with expiry time $T \geq 0$ and value $V$. Given a Wiener process $W_1(t)$, we model the change of the stock price with
\begin{equation}\label{eq:stock_change}
dS(t) = \mu S(t) dt + \sigma S(t) dW_1(t).
\end{equation}
While in the classical Black-Scholes model the volatility $\sigma$ is assumed to be constant, we follow \cite{HES} and represent it by yet another, correlated Wiener process $W_2(t)$. Then, denoting the volatility of volatility as $\xi$ we obtain the second stochastic differential equation
\begin{equation}\label{eq:var}
dv(t) =  \kappa (\gamma - v(t) ) dt + \xi \sqrt{v(t)}dW_2(t),
\end{equation}
where the variance $v=\sigma^{2}$ is a square of the volatility $\sigma$ and $\xi$ is assumed to be a known constant. We denote the correlation coefficient between $W_1$ and $W_2$ as $\rho \in (-1, 1)$. Note that \eqref{eq:var} is a mean-reverting process with a long term mean equal to $\gamma$ and a reversion level equal to $\kappa$. We take the drift coefficient $\mu$ in \eqref{eq:stock_change} to be constant and generally unknown.

At this point, we summarize the underlying assumptions of the model. The dividend payouts during the lifetime of the option are set to be $0$. We assume it is possible to lend and borrow any amount of a risk--free asset at a known constant interest rate $r$. Moreover, we are allowed to trade any amount, possibly fractional, of the stock $S$ or an option of value $V(t,S,v)$ at any time $0 \leq t \leq T$. We also say that the market is frictionless, which means that no such transaction generates fees. Lastly, we assume that there is no arbitrage opportunity.

It is well known \cite{HES}, also \cite[Chapter 51]{WIL}, that in this setting the existence of hedging portfolios ensures that the option price $V$ admits the parabolic equation
\begin{equation}\label{eq:hedging}
 \der{V}{t} + \frac{1}{2} v S^2 \sder{V}{S} + Sv\xi \rho \msder{V}{S}{v} + \frac{1}{2} v \xi^2 \sder{V}{v} + r S \der{V}{S} + (\kappa(\gamma -v) - \xi \lambda \sqrt{v})\der{V}{v} = 0,
\end{equation}
for some {\em universal} function $\lambda(S, \sigma, t)$ which common to all options. The function $\lambda$ is called the market price of volatility risk, because it can be interpreted as the value which market participants assign to the volatility risk. Yet, there is no agreed way to measure this function from market prices. Thus choosing the function $\lambda(S, \sigma, t)$ is extremely challenging from the theoretical point of view and as a practical task.

\cite{HES} proposes that it can be chosen equal to $\lambda \, \sigma$ with $\lambda \in \R$ being a scaling factor, i.e. in this special case $\lambda$ has the meaning of a coefficient and not the whole market price of volatility risk function. 

Since there is no agreed method of estimating the market price of volatility, one may argue that any such estimate will be burdened with inaccuracies. We propose therefore a new methodology to take this lack of knowledge about the market into account: We will model the market price of volatility risk as uncertain, borrowing concepts from \cite{ALP}, where the effect of uncertain volatility is examined.

More concretely, we assume that $\lambda$ is an unknown parameter contained in some interval $L \subset \R$ and consider the set $\mathbb{L}$ of all measurable mappings from $[0,T]$ to $L$. For all $\bm{\lambda} \in \mathbb{L}$ we define the linear operators $\mathcal{L}^{\bm{\lambda}}$ as
\begin{align}\label{eq:hestonop}
\textstyle - \frac{1}{2}\left( S^2 v \sder{V}{S} + 2 \rho \xi v S \msder{V}{S}{v} + \xi^2 v \sder{V}{v}\right) - r S \der{V}{S} - [\kappa (\gamma - v) - \xi \bm{\lambda}(t) \sqrt{v}\,] \der{V}{v} + r V.
\end{align}
We define the {\em Heston equation} associated to the control $\bm{\lambda} \in \mathbb{L}$ to be
\begin{equation}\label{eq:heston}
-\p_t V + \mathcal{L}^{\bm{\lambda}} V = 0. 
\end{equation}
Let us momentarily assume that $\bm{\lambda}$ is known and given. Then, in order to obtain a well-posed problem we need to also enforce boundary and the final time conditions. Throughout this chapter, given maturity $T$, we will use the following conditions
\begin{subequations} \label{eq:BC}
\begin{align}
    V(S,v,T) &= \Lambda(S), \label{eq:FinalTimeBC}  \\
    V(0,v,t) &= \Lambda(0), \label{eq:Dirichlet1BC}\\
    \lim_{S \to \infty} \der V S (S,v,t) &= \lim_{S \to \infty} \der \Lambda S(S), \label{eq:NeumannBC} \\
    -rS \der V S (S,0,t) - \kappa \gamma \der V v (S,0, t) \; + rV(S,0, t) - \partial_t V(S, 0, t) &= 0, \label{eq:nonlinearBC} \\
    \lim_{v \to \infty} \der V v (S,v,t) &= 0, \label{eq:Neumann2BC}
\end{align}
\end{subequations}
where the given function $\Lambda$ is the pay-off profile of the option. 

The boundary condition \eqref{eq:nonlinearBC} for a vanishing variance can be thought of as taking limit $v \to 0$ in the Black-Scholes equation and it is adapted directly from \cite{HES}. Notice on the other hand how compared to \cite{HES} the Dirichlet condition for a large volatility was replaced by the Neumann condition \eqref{eq:Neumann2BC}, which is more favourably from the numerical point of view. We motivate these Neumann conditions with the observation that as the volatility approaches extremely large values, influence of its oscillations on the option price is negligible. Thus we impose that the rate of change of $V$ in $v$-direction at this assymptotic boundary to be $0$. In the literature such approach was adopted for pricing of American options in \cite{American2} and \cite{largev_neumann2}, see also \cite{options_bc}.

\begin{remark} \label{rem:option_lambdas}
Recall that for a call option with the strike price $K$ one would choose $\Lambda$ as $\Lambda(S) = \max \left(0, S-K \right)$. In order to calculate the value of a long butterfly position of width $2a$ and the strike price $K$ the choice would be
\[
\Lambda(S) = \max\left( 0, S-(K-a) \right) -2 \max\left( 0, S-K \right) + \max\left( 0, S-(K+a) \right).
\]
Similarly, to consider the value of a long straddle with the strike price $K$ one would require $\Lambda(S) = \max\left( 0, S-K) \right) - \max\left( 0, S-K \right)$. \qed
\end{remark}

\begin{figure}
\begin{center}
  \begin{tikzpicture}[scale=1,auto]
    \draw (1,0) -- (4,0) -- (4,4) -- (1,4) -- (1,0);
    \draw [<->,thick] (0,4.5) node (zaxis) [above] {$v$} |- (4.5,0) node (yaxis) [right] {$S$};
    \draw[dashed] (1,4) -- (0,4);
	\node [left] at (-0.2,4) {$v_{\max}$};
	\node [left] at (-0.2,0) {$0$};
	\node [below] at (1,-0.2) {$S_{\min}$};
	\node [below] at (4,-0.2) {$S_{\max}$};
    \node at (2.5, 2) {$\O$};
	\node [below] at (2.5, 0) {$\pO_{R_t}$};
	\node [above] at (2.5, 4) {$\pO_{R_1}$};
	\node [left] at (1.0, 2) {$\pO_{D}$};
	\node [right]  at (4.0, 2) {$\pO_{R_2}$};
  \end{tikzpicture}
\end{center}
\caption{Truncated domain $\O = [S_{\min}, S_{\max}] \times [0, v_{\max}]$.}
\label{fig:heston_udomain}
\end{figure}
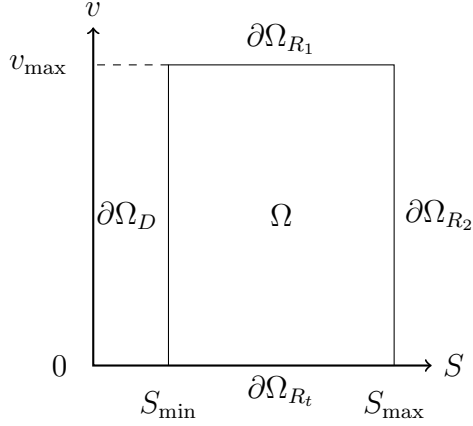

In practice, the implementation of the numerical scheme will require us to truncate the domain. Generally we choose a rectangular domain $\O = [S_{\min}, S_{\max}] \times [0, v_{\max}]$ as in Figure \ref{fig:heston_udomain} with
\begin{align*}
    \pO_{R_t} & = [S_{\min}, S_{\max}] \times \{0\},\\
    \pO_{R_1} & = [S_{\min}, S_{\max}] \times \{v_{\max}\},\\
    \pO_{R_2} & = \{ S_{\max} \} \times [0, v_{\max}],\\
    \pO_D & = \{ S_{\min} \} \times [0, v_{\max}].
\end{align*}
For the sake of brevity, we will also define $\Phi^{\bm{\lambda}}$ which we call the Heston operator:
\begin{align} \label{def:heston_op}
&\Phi^{\bm{\lambda}} V(t,S,v) = \\ \nonumber &\left\{ 
\begin{array}{rllll}
-\p_t V(t,S,v) + \mathcal{L}^{\bm{\lambda}(t)} V(t,S,v) &\textrm{if}\ (t,S,v) \in [0,T)\times \O,\\
V(t,S,v) - \Lambda(0) &\textrm{if}\ (t,S,v) \in [0,T) \times \pO_{D}, \\
(0, 1) \cdot \nabla V(t,S,v) &\textrm{if}\ (t,S,v) \in [0,T) \times \pO_{R_1}, \\
(1, 0) \cdot \nabla V(t,S,v) - \lim_{S \to \infty} \der \Lambda S (S) &\textrm{if}\ (t,S,v) \in [0,T) \times\pO_{R_2}, \\
-\p_t V(t,S,v) - (rS, \kappa \gamma) \cdot \nabla V(t,S,v) + rV(t,S,v) &\textrm{if}\ (t,S,v) \in [0,T)\times\pO_{R_t}, \\
V(t,S,v) - \Lambda(S) &\textrm{if}\  (t,S,v) \in \{T\}\times\oO.
\end{array}
\right.
\end{align}

\subsection{Extremal behaviour under uncertainty}

Of particular interest are the highest and lowest option price which can occur with market prices of volatility risk $\bm{\lambda} \in \mathbb{L}$. For $(t,S,v) \in [0,T] \times \overline{\Omega}$ we set
\begin{subequations} \label{eq:optimal_control}
\begin{align} 
\uV(t,S,v) &:= \inf_{\bm{\lambda} \in \mathbb{L}} \{ V^{\bm{\lambda}}(t,S,v) : \Phi^{\bm{\lambda}} V^{\bm{\lambda}} = 0 \text{ on } [0,T] \times \overline{\Omega} \},\\
\oV(t,S,v) &:= \sup_{\bm{\lambda} \in \mathbb{L}} \{ V^{\bm{\lambda}}(t,S,v) : \Phi^{\bm{\lambda}} V^{\bm{\lambda}} = 0 \text{ on } [0,T] \times \overline{\Omega} \}.
\end{align}
\end{subequations}
We seek an alternative characterisation of the functions $\uV$ and $\oV$ as solution of a PDE, more precisely of a Hamilton-Jacobi-Bellman (HJB) equation in combination with final time and boundary conditions. We shall focus here on the argument for $\uV$; the corresponding analysis for $\oV$ follows analogously. Notice also that the $\sup$ in the notation of $\uV$ refers to the $\sup$ in the below HJB equation \eqref{eq:HJB}, not to the $\sup$ on the right-hand side of \eqref{eq:optimal_control}.

Suppose there is a minimiser of the right-hand side of \eqref{eq:optimal_control}, which we denote $\bm{\hat{\lambda}}$. Then, for $(t,S,v) \in [0,T) \times \Omega)$ and $\uV$ sufficiently smooth, 
\begin{subequations} \label{eq:heston_dpp}
\begin{align}
    \uV(t,S,v)  & = \lim_{h \to 0} \Bigl[ \uV(t+h,S,v) - \int_t^{t+h} \partial_t \uV(\tau,S,v) d \tau \Bigr]\\
    & = \lim_{h \to 0} \Bigl[ \uV(t+h,S,v) - \int_t^{t+h} \mathcal{L}^{\bm{\hat{\lambda}}(t)} \uV(\tau,S,v) d \tau \Bigr]\\
    & = \lim_{h \to 0} \Bigl[ \uV(t+h,S,v) - \int_t^{t+h} \mathcal{L}^{\bm{\hat{\lambda}}(t)} \uV(\tau,S,v) d \tau \Bigr]\\
    & = \lim_{h \to 0} \Bigl[ \uV(t+h,S,v) - \int_t^{t+h} \mathcal{L}^{\bm{\hat{\lambda}}(t)} \uV(t + h,S,v) d \tau \Bigr]\\
    & = \lim_{h \to 0} \Bigl[ \uV(t+h,S,v) - \int_t^{t+h} \sup_{\lambda \in L} \mathcal{L}^{\lambda} \uV(t+h,S,v) d \tau \Bigr], \label{step:pointwise}
\end{align}
\end{subequations}
where \eqref{step:pointwise} follows from \eqref{eq:optimal_control}. After multiplication with $1 / h$ we find that
\begin{align*}
    \lim_{h \to 0} \frac{1}{h} \int_t^{t + h} \sup_{\lambda \in L} \mathcal{L}^{\lambda} \uV(\tau,S,v) d \tau & = \lim_{h \to 0} \frac{\uV(t + h,S,v) - \uV(t,S,v)}{h} \\
    & = \partial_t \uV(T,S,v).
\end{align*}
Assuming that $\uV$ is a classical solution so that
\begin{equation*}
    \lim_{h \to 0} \frac{1}{h} \int_t^{t + h} \sup_{\lambda \in L} \mathcal{L}^{\lambda} \uV(\tau,S,v) d \tau = \sup_{\lambda \in L} \mathcal{L}^{\lambda} \uV(t,S,v)
\end{equation*}
we arrive at
\begin{equation}\label{eq:HJB}
    -\partial_t \uV(t, S, V) + \sup_{\lambda \in L} \mathcal{L}^{\lambda} \uV(t, S, v) = 0.
\end{equation}
Because the boundary conditions \eqref{eq:BC} do not depend on $\lambda$, also $\uV$ satisfies them. Hence, in summary, $\uV$ solves the equation \eqref{eq:HJB} subject to the boundary conditions~\eqref{eq:BC}. 

\section{Transformation of the uncertain Heston model} \label{sec:trans}

We propose to use the finite element method of \cite{mixedBC} for the numerical approximation of $\uV$. It is shown there that this method can capture the fully nonlinear structure of \eqref{eq:HJB} as well as the mixed (and thus discontinuous) boundary conditions \eqref{eq:BC} and that it will converge under mesh refinement to the viscosity solution of the final time boundary value problem. Importantly, this finite element approach has been shown \cite{IJEN} to guarantee strong convergence in the gradient of the value function, even for a degenerately elliptic HJB operator, as is the case here. 

\begin{figure}
\begin{center}
  \begin{tikzpicture}[scale=1,auto]
    \draw (1,0) -- (4,0) -- (4,4) -- (1,4) -- (1,0);
    \draw [<->,thick] (0,4.5) node (zaxis) [above] {$v$} |- (4.5,0) node (yaxis) [right] {$S$};
    \draw[dashed] (1,4) -- (0,4);
	\node [left] at (-0.2,4) {$v_{\max}$};
	\node [left] at (-0.2,0) {$0$};
	\node [below] at (1,-0.2) {$\log(S_{\min})$};
	\node [below] at (4,-0.2) {$\log(S_{\max})$};
    \node at (2.5, 2) {$\O'$};
	\node [below] at (2.5, 0) {$\pO'_{R_t}$};
	\node [above] at (2.5, 4) {$\pO'_{R_1}$};
	\node [left] at (1.0, 2) {$\pO'_{D}$};
	\node [right]  at (4.0, 2) {$\pO'_{R_2}$};
  \end{tikzpicture}
\end{center}
\caption{Domain $\O^{'}$ after the transformation $S=e^x$}
\label{fig:heston_1domain}
\end{figure}
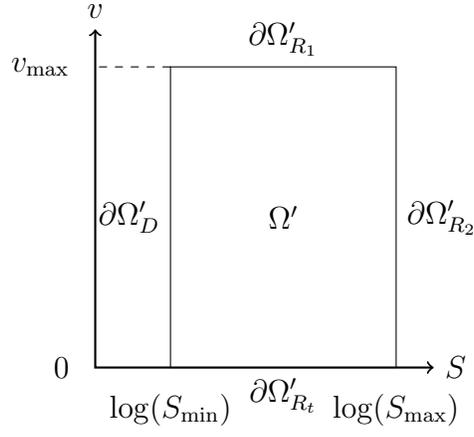

In this section will now perform the transformation of the elliptic operators $\mathcal{L}^{\lambda}$ to their isotropic form in order to be consistent with the framework of the numerical method formulated in Chapter \cite{mixedBC}. Our first goal is to remove the $S$ dependence of the coefficients. In order to do that we let $S = e^{x}$. Then
\begin{equation*}
\der{\uV}{x} = S \der{\uV}{S},\quad
\sder{\uV}{x} = S^2 \sder{\uV}{S}+ \der{\uV}{x}, \quad
 \msder{\uV}{S}{v} = \msder{\uV}{x}{v}
\end{equation*}
where the transformed domain for $x$ is denoted $\Omega'$, see Figure \ref{fig:heston_1domain}. Substituting into \eqref{eq:hestonop} we get
\begin{align}\label{eq:lnheston}
\mathcal{L}_{1}^{\lambda}\uV(x,v,t) :=& - \frac{1}{2}v\left( \sder{\uV}{x} + 2\rho\xi \msder{\uV}{x}{v} + \xi^2 \sder{\uV}{v}\right) \nonumber \\
&-(r-\frac{1}{2}v)\der{\uV}{x} -  [\kappa (\gamma - v)-\xi \lambda \sqrt{v}\,] \der{\uV}{v} + r\uV.
\end{align}
Similarly, with
\begin{equation*}
 \der{\uV}{S} = \der{\uV}{x}\der{x}{S} = e^{-x}\der{\uV}{x}, \quad \der{\Lambda}{S} = \der{\Lambda}{x}\der{x}{S} = e^{-x}\der{\Lambda}{x}
\end{equation*}
the transformed Neumann boundary condition \eqref{eq:NeumannBC} is
\begin{equation}\label{ln_Neumann}
   \der{\uV(x,v,t)}{x} \Bigr|_{(x,v) \in \pO^{'}_{R_2}} = \lim_{x \to \infty} \der{\Lambda(S(x))}{x}.    
\end{equation}
The boundary conditions on $\pO'_{D}$ and $\pO'_{R_1}$ remain as in \eqref{eq:Dirichlet1BC} and \eqref{eq:Neumann2BC}, in the final time condition one substitutes $e^x$ for $S$ and the Robin condition on $\pO_{R_t}$ is obtained by substituting $v=0$ into \eqref{eq:lnheston}. The Heston operator $\Phi_1$ in the new coordinates is then defined by
\begin{align} \label{def:heston1_op}
& \Phi^{\bm{\lambda}}_{1} V(t,x,v) =   \\ & \left\{ 
\begin{array}{rllll}
-\p_t V(t,x,v) + \mathcal{L}^{\bm{\lambda}(t)}_{1} V(t,x,v) &\textrm{if}\ (t,x,v) \in[0,T)\times \O^{'},\\
V(t,x,v) - \Lambda(0) &\textrm{if}\ (t,x,v) \in[0,T) \times \pO^{'}_{D}, \\
(0, 1) \cdot \nabla V(t,x,v) &\textrm{if}\ (t,x,v) \in[0,T) \times \pO^{'}_{R_1}, \\
(1, 0) \cdot \nabla V(t,x,v) - \lim_{x \to \infty} \der{\Lambda(S(x))}{x}&\textrm{if}\ (t,x,v) \in[0,T) \times\pO^{'}_{R_2}, \\
-\p_t V(t,x,v) - (r, \kappa \gamma) \cdot \nabla V(t,x,v) + rV(t,x,v) &\textrm{if}\ (t,x,v) \in [0,T)\times\pO^{'}_{R_t}, \\
V(t,x,v) - \Lambda(S(x)) \; &\textrm{if}\ (t,x,v) \in \{T\}\times\overline{\O^{'}}. \nonumber
\end{array}
\right.
\end{align}

In order to remove the second order mixed derivative from $\mathcal{L}_1^{\lambda}$ we consider the following change of variables:
\begin{equation*}
y = x - \frac{\rho}{\xi}v, \quad
z = \frac{\sqrt{1-\rho^2}}{\xi}v,
\end{equation*}
where $y \in \mathbb{R}$ and $z\geq 0$. The domain $\O^{'}$ is transformed into $\O^{''}$ whose shape in general depends on the parameters of the numerical experiment. It is depicted in Figure \ref{fig:heston_1domain} with the numerical values of the case study in Section \ref{sec:case}.

\begin{figure}
     \centering
  \begin{tikzpicture}[scale=1,auto]
\draw (0,0) -- (4,0) -- (2,4) -- (-2,4) -- (0,0);
\draw [<->,thick] (0,4.5) node (zaxis) [above] {$z$} |- (4.5,0) node (yaxis) [below] {$y$};
	  \node [left] at (-2,4) {$(-2.14,\;3.7)$};
	  \node [right] at (2,4) {$(2.46,\;3.7)$};
	  \node [left] at (-0.2,0.3) {$(0,\;0)$};
	  \node [right] at (4, 0.3) {$(4.61,\;0)$};
      \node at (1, 2) {$\O^{''}$};
	  \node [below] at (2, 0) {$\pO^{''}_{R_t}$};
	  \node [above] at (0.5, 4) {$\pO^{''}_{R_1}$};
	  \node [left] at (-1.1, 2) {$\pO^{''}_{D}$};
	  \node [right]  at (3.1, 2) {$\pO^{''}_{R_2}$};
	  \end{tikzpicture}
         \caption{Domain $\O^{''}$ with parameter values $\rho=0.5$, $\xi=0.7$}
        \label{fig:heston_2domain}
\end{figure}
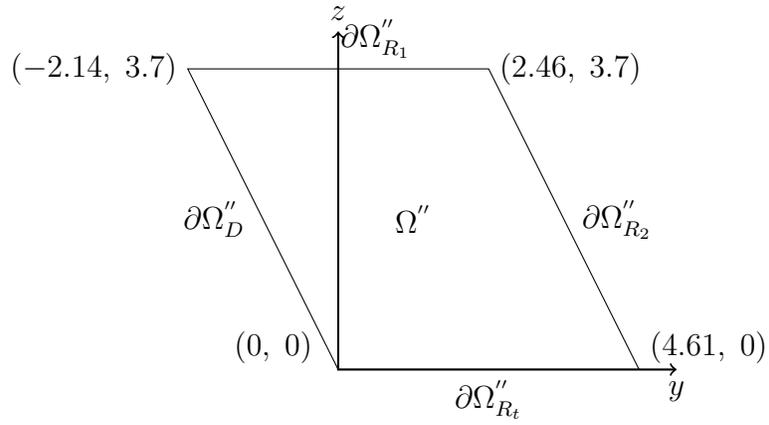
For $w(y,\,z,\,t) := \uV\left(x(y,\,z),\,v(y,\,z),\,t\right)$ we have that
\begin{equation*}
\begin{aligned}[t]
\tst \der{\uV}{x} &\tst= \der{w}{y},\\
\tst\der{\uV}{v} &\tst= -\frac{\rho}{\xi} \der{w}{y} + \frac{\sqrt{1-\rho^2}}{\xi}\der{w}{z},\\
\tst\sder{\uV}{x} &\tst= \sder{w}{y}. 
\end{aligned}
\qquad
\begin{aligned}[t]
\tst\sder{\uV}{v} &\tst= \frac{\rho^2}{\xi^2} \sder{w}{y} - \frac{2 \rho \sqrt{1-\rho^2}}{\xi^2} \msder{w}{y}{z} + \frac{1-\rho^2}{\xi^2}\sder{w}{z},\\
\tst\msder{\uV}{x}{v} &\tst= -\frac{\rho}{\xi}\sder{w}{y} + \frac{\sqrt{1-\rho^2}}{\xi} \msder{w}{y}{z},\\
\end{aligned}
\end{equation*}
Combining those results with \eqref{eq:lnheston} we obtain the canonical formulation of every $\mathcal{L}^{\bm{\lambda}}$, $\bm{\lambda} \in \mathbb{L}$, from \eqref{eq:heston} as required:
\begin{align}\label{eq:canop}
\mathcal{L}^{\bm{\lambda}}_{2} w := \, & \frac{-\xi\sqrt{1-\rho^2}}{2}z \Delta w + \left(-r + \frac{\kappa \gamma \rho}{\xi}+\frac{\frac{1}{2}\xi -\kappa\rho}{\sqrt{1-\rho^2}}z  - \bm{\lambda} \rho \sqrt{\frac{\xi z}{\sqrt{1-\rho^2}}}\right)\der{w}{y} \nonumber\\
&+ \left( \frac{-\kappa \gamma \sqrt{1-\rho^2}}{\xi} + \kappa z + \bm{\lambda} \sqrt{\xi z \sqrt{1-\rho^2}}\right)\der{w}{z} + rw. 
\end{align}
Now we reformulate \eqref{eq:FinalTimeBC}-\eqref{eq:Neumann2BC} accordingly. Since $\der \uV x = \der w y$, the Neumann boundary condition \eqref{eq:NeumannBC} for large stock prices is obtained by simply substituting $y$ and $z$ into \eqref{def:heston1_op} on $[0,T) \times \pO^{'}_{R_2}$ which results in
 \begin{equation}\label{t_Neumann}
    \der{w}{y} \Bigr|_{(y,z) \in \pO^{''}_{R_2}} = \lim_{y \to \infty} \der{\Lambda(S(x(y,z)))}{y}    
 \end{equation}
Under the aforementioned change of variables the Dirichlet boundary condition \eqref{eq:Dirichlet1BC} noticing that $\lim_{S \to 0}y = -\infty$ converts to
\[
w(y,\,z,\,t) \Bigr|_{(y,z) \in \pO^{''}_D}= \Lambda(0),
\]
while for the Neumann condition \eqref{eq:Neumann2BC} we use the fact that $\der{\uV}{v} = -\frac{\rho}{\xi} \der{w}{y} + \frac{\sqrt{1-\rho^2}}{\xi}\der{w}{z}$ and $\lim_{v \to \infty}z = \infty$ to obtain
\[
        \left(\frac{-\rho}{\xi}, \frac{\sqrt{1-\rho^2}}{\xi} \right) \cdot \nabla w= 0.
\]
Analogously to the result in \cite{HES}, the Robin boundary condition for $v \to 0$ is obtained simply by substituting $z=0$ into \eqref{eq:canop} which gives
\begin{equation}\label{t_Robin}
-\p_t w + \left(-r + \frac{\kappa \gamma \rho}{\xi}\right) \der{w}{y}  +
\left( \frac{-\kappa \gamma  \sqrt{1 - \rho^2}}{\xi} \right)\der{w}{z} + rw =0. 
\end{equation}
We summarize the above results by introducing the transformed Heston operator $\Phi_{2}$ defined as follows
\begin{align} \label{def:heston2_op}
& \Phi^{\bm{\lambda}}_{2}w (t,y,z) =  \\ & \left\{ 
\begin{array}{rllll}
-\p_t w(t,y,z) + \mathcal{L}^{\bm{\lambda}(t)}_{2} w(t,y,z) &\textrm{if}\ (t,y,z) \in[0,T)\times \O^{''},\\
w(t,y,z) - \Lambda(0) &\textrm{if}\ (t,y,z) \in[0,T) \times \pO^{''}_{D}, \\
(\frac{-\rho}{\xi}, \frac{\sqrt{1-\rho^2}}{\xi}) \cdot \nabla w(t,y,z) &\textrm{if}\ (t,y,z) \in[0,T) \times \pO^{''}_{R_1}, \\
(1, 0) \cdot \nabla w(t,y,z) - \lim_{y \to \infty} \der{\Lambda(S(x(y,z)))}{y}&\textrm{if}\ (t,y,z) \in[0,T) \times\pO^{''}_{R_2}, \\
\hspace*{-2mm}-\p_t w(t,y,z) - (r + \frac{\kappa \gamma \rho}{\xi}, \frac{-\kappa \gamma  \sqrt{1 - \rho^2}}{\xi}) \cdot \nabla \uV 
+ rw(t,y,z) &\textrm{if}\ (t,y,z) \in [0,T)\times\pO^{''}_{R_t}, \\
w(t,y,z) - \Lambda(S(x(y,z))) \; &\textrm{if}\ (t,y,z) \in \{T\}\times\overline{\O^{''}}. \nonumber
\end{array}
\right.
\end{align}
By replacing the Heston operator $\Phi^{\bm{\lambda}}$ from \eqref{eq:hestonop} with its transformed version from \eqref{def:heston2_op} and following the same argument as in the previous section we obtain an optimal control problem 
\begin{align} \label{eq:transformedHJB}
\sup_{\lambda \in L} \Phi^{\lambda}_{2} w(t,y,z) = 0 \qquad \forall (t,y,z) \in [0,T] \times \overline{\Omega^{''}} 
\end{align}
analogous to \eqref{eq:BC} and \eqref{eq:HJB} with the structure conforming to the setting of \cite{mixedBC}. Note that it resembles the "worst-case scenario" described in \cite{ALP} but with $\lambda$ instead of $\sigma$ taking the role of the uncertain parameter. Having completed the transformation, this allows us to treat the market price of volatility risk as a control in an isotropic HJB problem.

\section{Case Study} \label{sec:case}

We now investigate the qualitative and quantitative effects of the market price of volatility risk on the price of an option. The computations use the finite element discretisation of \cite{mixedBC}. The code is available from the public repository \cite{github} under the GNU Lesser General Public License.

As benchmark problem we consider the parameters of the experiment described in \cite[Table 11]{Doran_thesis}. Following this setting we choose the final time $T=0.5$, the strike price $K=50$, the volatility of volatility $\xi = 0.7$, the long term volatility mean $\gamma=0.3$, the mean reversion rate $\kappa=7$ and the correlation parameter $\rho=0.5$. Since the risk free rate $r$ is not stated explicitly, we made a choice of $r=0.03$. The domain is truncated with $v \in [0,3]$ and $S \in [1,100]$, which results in the transformed domain seen in Figure \ref{fig:heston_2domain}. In the source \cite{Doran_thesis} a simple call option is studied, which for example with $L=[-2.4, -1.6]$ has the constant optimal control $\hat{\bm{\lambda}} \equiv -2.4$. Instead we turn our attention here to a long butterfly position of width $40$ which, as mentioned in Remark \ref{rem:option_lambdas}, is equivalent to choosing
\[
    \Lambda(S) = \max\left( 0, S-30 \right) -2 \max\left( 0, S-50 \right) + \max\left( 0, S-70 \right).
\]

\subsection{Result 1: Value Function}

\begin{figure}
\centering
\includegraphics[width=\textwidth]{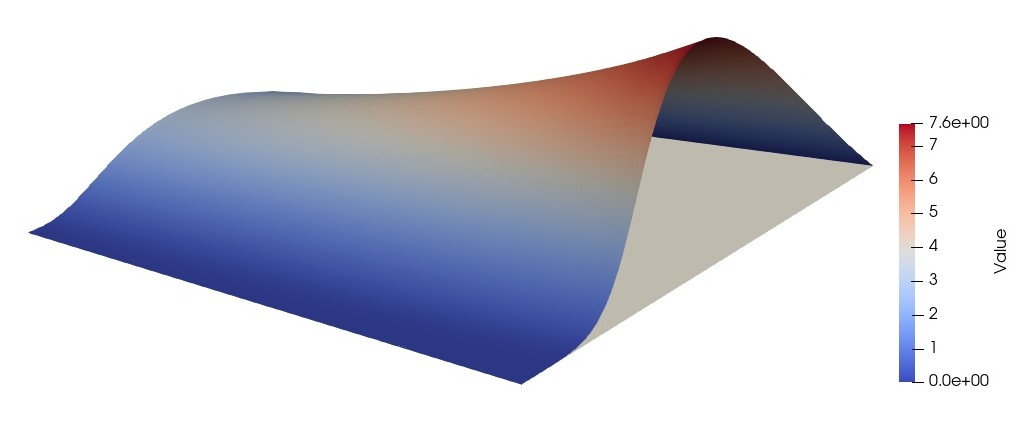}
\caption{Value of a long butterfly position at $t=0$ with $T=0.5$, $K=50$ and control set $L=[-2.4, -1.6]$.}
\label{fig:butterfly_value_func}
\end{figure}

\begin{figure}
\begin{subfigure}[t]{.45\textwidth}
  \centering
  \includegraphics[width=\linewidth]{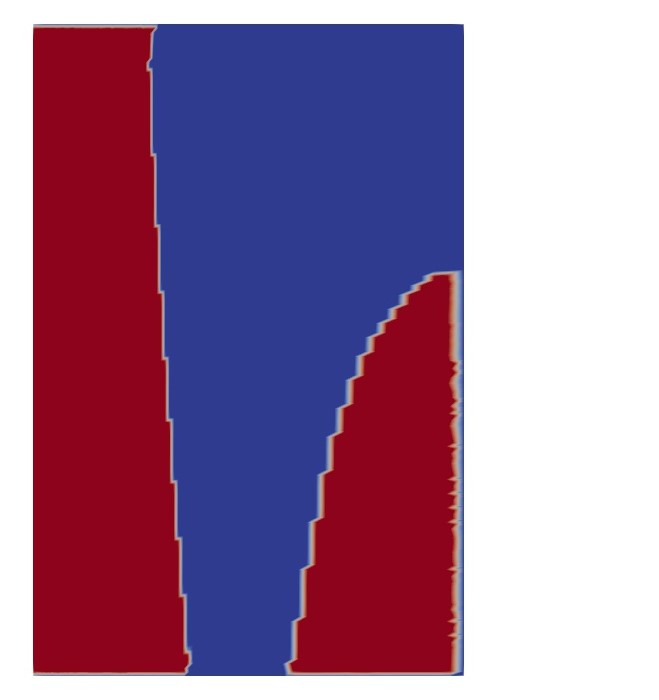}
  \caption{Selected optimal control; which are throughout at the extreme points of the control set.}
  \label{fig:butterfly_control}
\end{subfigure}%
\hspace{2em}
\begin{subfigure}[t]{.45\textwidth}
  \centering
  \includegraphics[width=\linewidth]{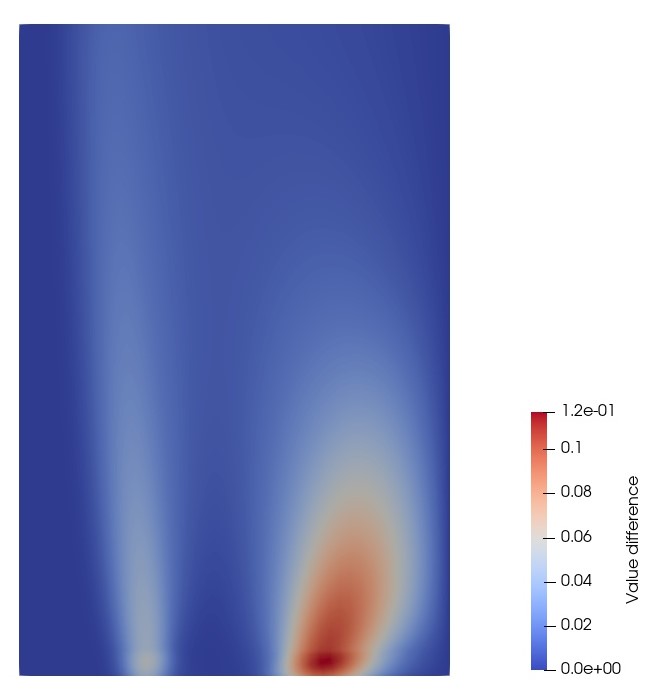}
  \caption{Difference between solutions of a nonlinear problem and linear evolution problem with a fixed control $\lambda = -2.4$.}
  \label{fig:linear_vs_nonlinear}
\end{subfigure}
\caption{Measurement of the effect of non-linearity for a long butterfly position at $t \approx 0.39$ with $T=0.5$, $K=50$ and control set $L=[-2.4, -1.6]$.}
\label{fig:control_testing}
\end{figure}

We let $L=[-2.4, -1.6]$. Note how interval $L$ is centred around the market price of volatility risk equal to $-2$ used in \cite{Doran_thesis}. The numerical approximation of the solution to the HJB problem is performed on the transformed domain $\O^{''}$ and then the resulting function is cast back to original domain $\O$. The outcome is depicted in Figure \ref{fig:butterfly_value_func}. Moreover, one can see in Figure~\ref{fig:butterfly_control} that the numerical method in fact selects different controls as optimal in different areas of the domain. The difference between the solution of the nonlinear problem compared to the solution of the linear evolution problem associated to one of the controls can be seen in Figure \ref{fig:linear_vs_nonlinear}. This highlights the importance of using a nonlinear model.

\subsection{Result 2: $\lambda$ interval testing}

\begin{figure}
\centering
\subfloat[Comparison of $\uV$ and $\oV$ at $(S,v)=(2.11,2.06)$ ]{\label{fig:intervals_value}\includegraphics[width=0.45\linewidth]{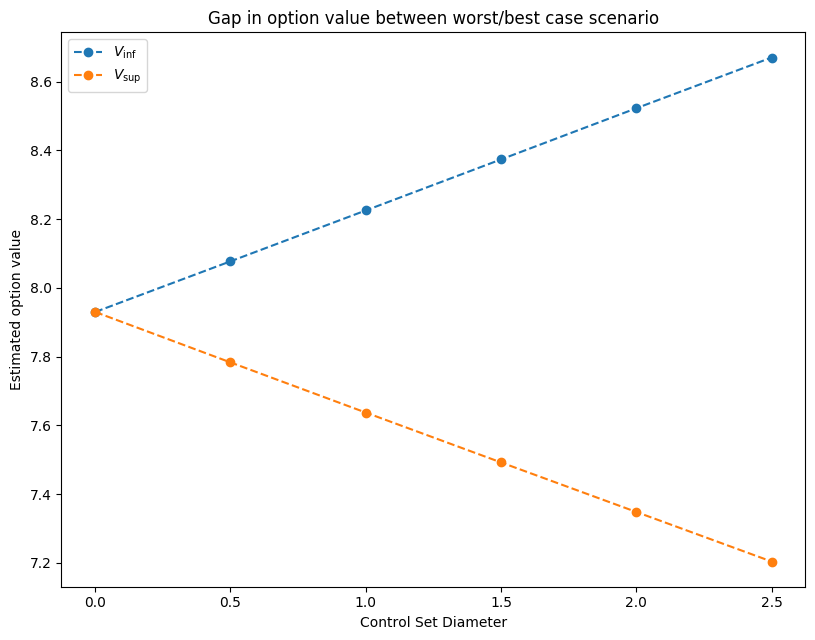}}\qquad
\subfloat[Comparison of $\partial \uV/\partial S$ and $\partial \oV/\partial S$ at $(S,v)=(53.12,0.75)$]{\label{fig:intervals_deltasb}\includegraphics[width=0.45\linewidth]{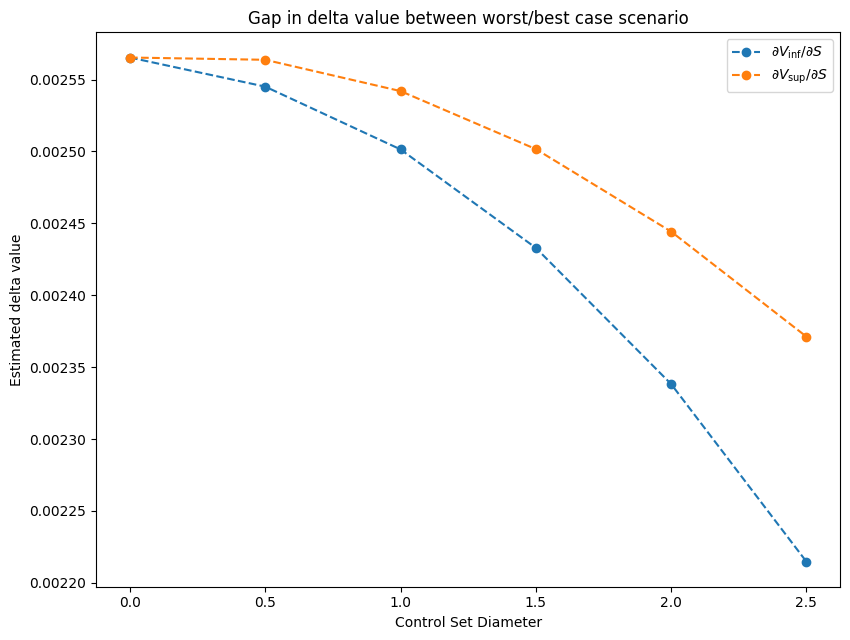}}\\
\subfloat[Comparison of $\partial \uV/\partial S$ and $\partial \oV/\partial S$ at $(S,v)=(51.76,2.84)$]{\label{fig:intervals_deltasc}\includegraphics[width=0.45\textwidth]{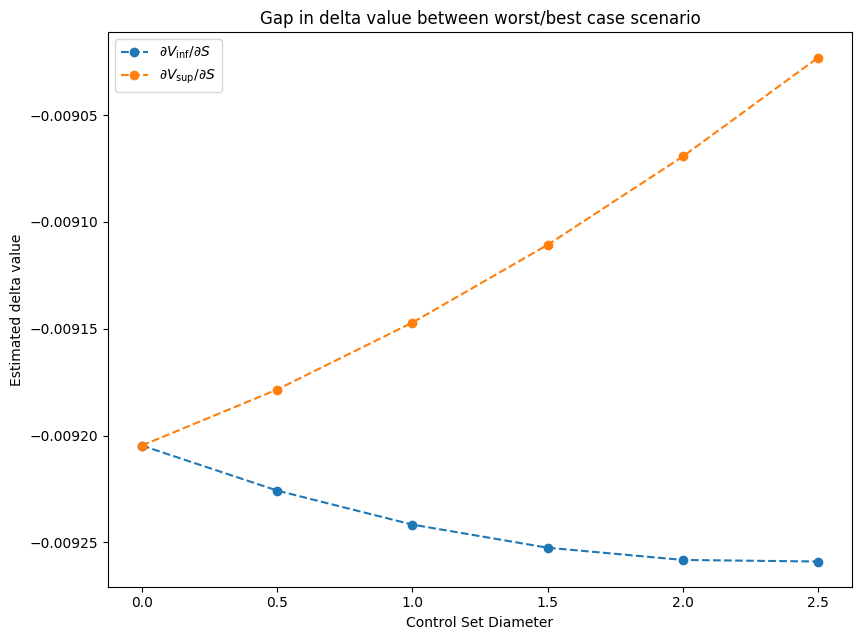}}\qquad%
\subfloat[Comparison of $\partial \uV/\partial S$ and $\partial \oV/\partial S$ at $(S,v)=(51.43,0.23)$]{\label{fig:intervals_deltasd}\includegraphics[width=0.45\textwidth]{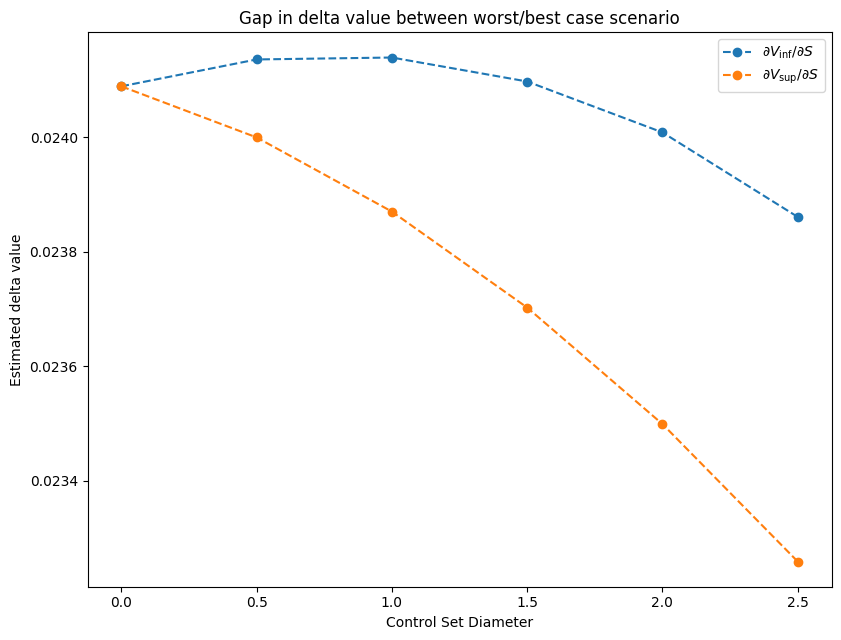}}%
\caption{Measurement of effect of a diameter of a control set on the value function and its derivative. Control sets are symmetrical and centred at $-1.25$, measurements were made at $t=0$}
\label{fig:intervals}
\end{figure}

We now assess the impact of different choices of control sets $L$ on the option value estimate. Indeed, we consider control sets of increasing diameter and measure the difference between the value function $\oV$ of the worst case scenario and the value function $\uV$ of the best case scenario. The results are shown in Figure \ref{fig:intervals}. The computations show the significant effect of the uncertainty in the market price of volatility risk on the option price. As indicated by Figure \ref{fig:intervals_value} the option value of worse and best case scenario can differ up to $16\%$. Note that in this case the control set contains values ranging between $0$ and $-2.5$, which were found to be used in the literature. Given the evidence (see for example findings in \cite{BAK}) that market price of volatility takes negative values, the simplification of taking $\lambda=0$ may lead to erroneous estimates. On the other hand, the experiments indicate a linear correlation and in general more negative market prices of volatility risk lead to higher option values. 

We now direct our attention to the partial derivatives of option value $V$ since they are used to create hedging portfolios. We investigate the effect of $\lambda$ on the partial derivative of the option value with respect to $S$. As seen in Figures \ref{fig:intervals_deltasb}-\ref{fig:intervals_deltasd} the impact of the value of $\lambda$ on Delta $\partial V/\partial S$ is strongly nonlinear in the vicinity of the strike price $K$. We remark at this point that numerical methods which do not guarantee gradient convergence may in general fail to capture this kind of behaviour.

Figure \ref{fig:intervals} provides a visual portrayal of the sensitivity of the price and Delta on the magnitude of the uncertainty.

\subsection{Result 3: Delta plots}

\begin{figure}
\begin{subfigure}{.5\textwidth}
\centering
  \includegraphics[width=.8\linewidth]{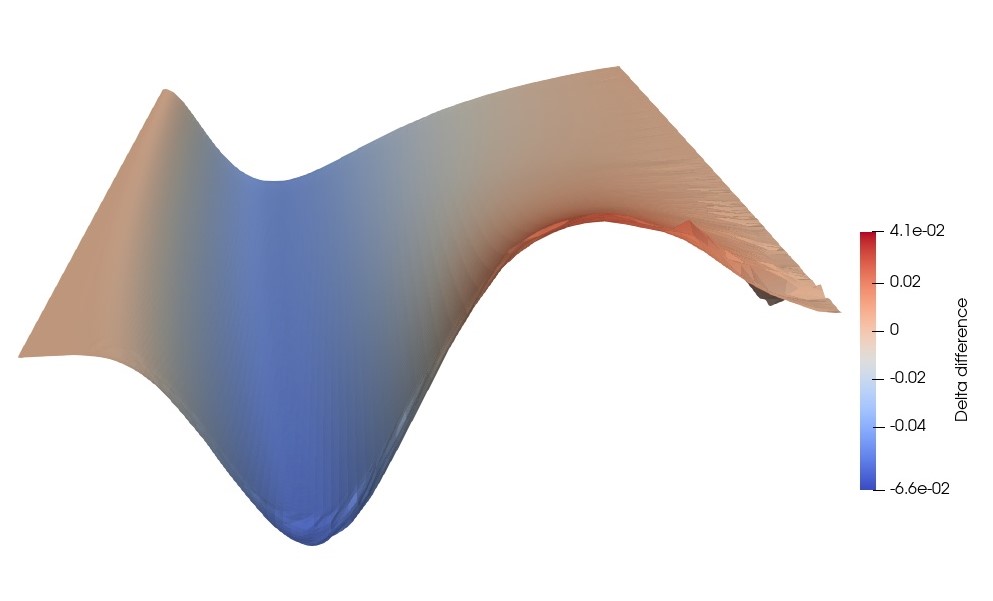}
  \caption{Call option}
  \label{fig:call_deltaS}
\end{subfigure}%
\begin{subfigure}{.5\textwidth}
  \centering
  \includegraphics[width=.8\linewidth]{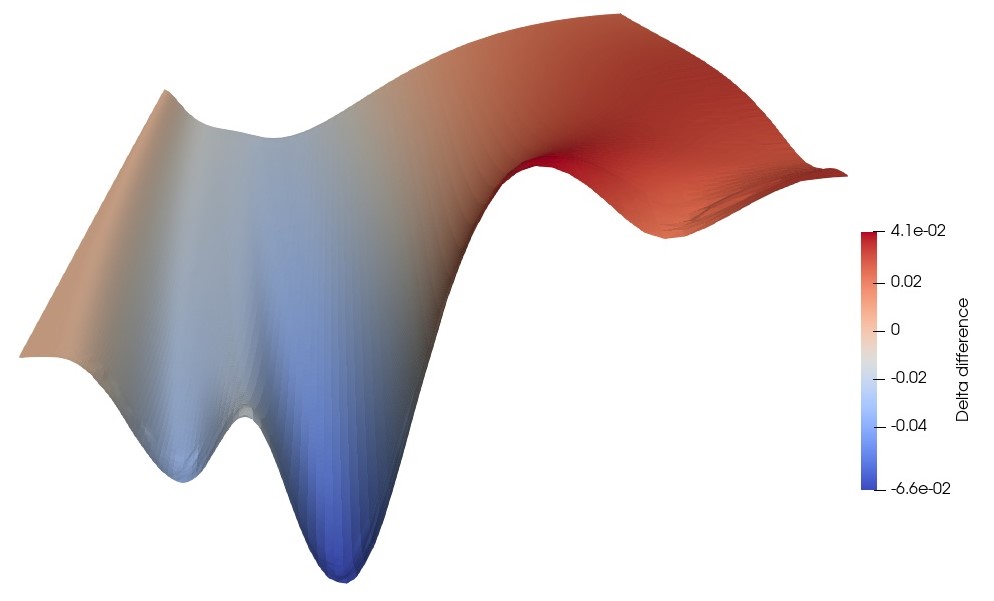}
  \caption{Butterfly option}
  \label{fig:butterfly_deltaS}
\end{subfigure}
\caption{Comparison of plots of $\delta(\uV-\oV)/\delta S$ at time $t=0$ with control set $[-2.5,0.0]$}
\label{fig:call_delta_plots}
\end{figure}

In line with the results of the previous experiment, we continue to investigate the worst and the best case scenarios for the control set $L=[-2.5,0.0]$ at time $t=0$. We plot differences between the Deltas $\partial \uV/\partial S$ and $\partial \oV/\partial S$ for all points in $\O$ at time $t=0$. The results for a call option are shown in Figure \ref{fig:call_deltaS} and for a long butterfly option in Figure \ref{fig:butterfly_deltaS}. Note that since $\partial \oV/\partial S$ and $\partial \uV/\partial S$ are both of order $1$, the graphs represent a relative as well as an absolute error. We conclude that the impact of the market price of volatility risk on the delta values is significant. In the covered examples, one can expect up to $6\%$ difference between the scenario where $\lambda$ is neglected and the one where the HJB approach is used.

\section*{Funding}

Bartosz Jaroszkowski acknowledges the support of the EPSRC grant 1816514. Max Jensen acknowledges the support of the Dr Perry James Browne Research Centre.


\bibliographystyle{alpha}
\bibliography{references}


\end{document}